\documentclass[twocolumn,preprintnumbers, amsmath, amssymb, floatfix]{revtex4}
\usepackage{dcolumn}
\usepackage{bm}
\def\be{\begin{equation}}
\def\ee{\end{equation}}
\def\beq{\begin{eqnarray}}
\def\eeq{\end{eqnarray}}
\def\n{\nonumber}
\def\bay{\begin{array}}
\def\eay{\end{array}}

\def\morphf{{\raisebox{-0.28ex}{$\mathrm{f}$} \atop \raisebox{0.8ex}{$\to$}}}
\def\isomorph{{\raisebox{-0.28ex}{$\sim$} \atop \raisebox{0.8ex}{$\rightarrow$}}}
\def\dashedmorph{{\raisebox{-0.28ex}{$\mathrm{e}$} \atop \raisebox{0.8ex}{$\dashrightarrow$}}}
\def\morphupdown{ {{\raisebox{-0.28ex} {$\mathrm{p}$} \atop
\raisebox{0.8ex} {$\rightrightarrows$}} \atop {\raisebox{-0.28ex}
{$\mathrm{q}$}}} }
\begin{document}
\preprint{smw05-sams-conf02}
\title{Universal Relativity and Its Mathematical Requirements}

\author{Sanjay M. Wagh}
\altaffiliation[Permanent Address:]{Central India Research
Institute, Post Box 606, Laxminagar, Nagpur 440 022, India
\\Electronic address: {\bf\tt
cirinag\underline{\phantom{n}}ngp@sancharnet.in}}
\affiliation{School of Mathematical Sciences, University of
KwaZulu-Natal, Durban 4041, Republic of South Africa}

\email{waghs@ukzn.ac.za}

\date{December 8, 2005}
\begin{abstract}
In this presentation, I review physical principles behind a
recently proposed \cite{smw-utr} Universal Theory of Relativity
and speculate on the mathematical requirements implied by these
physical principles. Some unresolved issues will also be discussed. \\

\centerline{Presented at the South African Mathematical Society's
48th Meeting} \centerline{October 31 - November 2, 2005, Rhodes
University, Grahamstown, South Africa}
\end{abstract}
\maketitle

\newpage
Newtonian explanations of physical phenomena rest on four, {\em
mutually logically independent}, conceptions, namely, those of
inertia, force, source of the force and a law of motion. Due to
the mutual logical independencies of these conceptions, these are
the {\em assumptions or postulates\/} of the newtonian theoretical
framework. The same logical independency of basic concepts is also
behind ``the equality of inertia and gravitational mass of a
physical body'' being an assumption of Newton's theory. Thus,
newtonian theories cannot offer explanations for the inertia, the
source characteristics of physical matter, the law of force and
the law of motion. Within these limitations, newtonian theories
still explain a large body of observations.

To explain the known experimental data, we then need to assume
{\em four\/} fundamental forces, {\em viz}, those of gravity,
electromagnetism, strong nuclear and weak nuclear interactions.

Theoretical physicists, in general, aim to ``unify'' these four
basic forces. That is to say, the grand aim of theoretical
physicists is to `show' that these fundamental forces arise from a
single, mathematical or physical, entity. This ``unification of
the four (fundamental) forces'' is then required \cite{smw-utr} to
also provide us an ``explanation'' that removes the mutual logical
independencies of the involved fundamental physical concepts.
Evidently, this aim requires departures from the newtonian
conceptual as well as mathematical framework.

Although, for some while, newtonian explanations were considered
to be encompassing the entirety of observable phenomena, the
experimental data forced \cite{schlipp} departures away from the
newtonian conceptual framework. This happened, as is well known,
towards the end of the 19th century. Subsequent theoretical works
progressed along the following two separate directions.

Firstly, Special Relativity expanded the Galilean group to the
Lorentz group of transformations (of the same coordinates used by
Newton to express his Laws of Motion). This ``group enlargement''
provided explanations for certain physical phenomena, in
particular, those from electromagnetism. These explanations were
considered ``natural'' because Maxwell's equations, describing
this radiation reasonably successfully, are form-invariant under
the Lorentz group.

The same group enlargement also implies ``modifications'' to the
Newtonian laws of motion and leads to ``theoretical relations''
such as the inertia-energy equivalence. These led to experimental
predictions and to explanations of further physical phenomena
lending credence to those theoretical relations. Einstein,
specifically, attached \cite{schlipp} importance to the special
relativistic result of the variation of inertia with velocity.
That only the group enlargement leads to this result left an
indelible mark on Einstein's subsequent works which led him to the
Principle of General Relativity which we shall consider at a later
stage.

Black body radiation, photoelectric effect, spectroscopy forced
the second theoretical direction. Such physical phenomena led to
mathematical techniques away from the standard newtonian and
special relativistic ones. These, quantum theoretic, techniques
built on the newtonian notions using Hilbert spaces, and led to
specific experimental predictions. For example, Quantum Theory
predicted the phenomenon of tunnelling across a barrier, and it
was observed. Classical methods had no explanations for this as
well as many other observed phenomena.

However, methods of neither special relativity \footnote{Special
relativity and Newton's theory, both, remain silent on as to how
to treat the physical laws in relation to systems of reference in
acceleration relative to inertial systems. It therefore needs to
postulate pseudo-forces in laws stated relative to non-inertial
systems. There are no `sources' in material bodies for the
pseudo-forces and, hence, these represent `unnatural' or
`extraneous' quantities that we have to postulate \cite{smw-utr}.}
nor can quantum theory \footnote{Quantum theory replaces the
notion of force by the corresponding potential and uses
mathematical techniques (involving Hilbert spaces) different than
the standard newtonian ones. Nonetheless, quantum theory is based
on inertia, potential, source of the potential and a law of
potential as mutually logically independent concepts. Methods of
quantum theory cannot then lead to `unification' of the four
fundamental physical interactions in the sense as described in the
sequel.} claim to remove the mutual logical independencies of
fundamental newtonian notions. Rather, these methods, implicitly
or explicitly, replace these notions by other equivalent concepts
which too are mutually logically independent. Thence, these
methods are inadequate to achieve the aim of the unification of
the four fundamental forces.

Einstein then took the bold step of formulating judiciously
constructed {\em Gedanken\/} or thought experiments and arrived at
his Principle of General Relativity that the Laws of Physics
should be applicable with respect to all the systems of reference,
in relative acceleration or not, without extraneous or unnatural
entities (like pseudo-forces) serendipitously entering into them.

Einstein's experience with the special relativistic group
enlargement led him to realize that ``group enlargement'', further
than that of the Lorentz group of Special Relativity, is demanded
by this Principle of General Relativity. We then note that Laws of
Physics are some mathematical statements about mathematical
structures representing physical or ``observable'' bodies.

Clearly, there are two issues involved here: first that of
deciding which mathematical structures are to represent physical
bodies and, second, that of mathematical statements (equations)
about these structures corresponding to physical or observable
changes imagined as the transformations of these mathematical
structures.

Einstein, in recognition of the above, wrote \cite{schlipp}:
\medskip

$\spadesuit$ {\em Of which mathematical type are the variables
(functions of the coordinates) which permit the expression of the
physical properties of the space (``structure'')? Only after that:
Which equations are satisfied by those variables?} $\spadesuit$
\medskip

It is crucial to recognize here that the systems of reference in
physics are `constructed' out of the very physical matter or
bodies which are the subject of physical laws. That is to say, a
physical body gets used as a reference when physical measurements
are performed. Then, the very physical phenomena that are the
subject of study in Physics {\em affect\/} the physical
construction of the reference systems. Hence, the validity of
Einstein's principle of general relativity is incorporating,
implicitly or explicitly, the physical changes that can occur to
reference systems when various physical phenomena take place.

Consequently, only that mathematical framework which allows the
physical systems of reference to be `affected' by physical
phenomena can be consistent with the general principle of
relativity. {\em This situation essentially provides us a {\bf
guiding principle} for `selecting' mathematical structures to
represent physical bodies}.

To illustrate the use of this guiding principle, let us consider
the geometry of the surface of a sphere. Mathematical coordination
of the points of the spherical surface is effected by choosing the
coordinate axes as the equator and a great circle passing through
the north and the south poles. Further mathematical apparatus is
then developed in the usual way by following the methodology of
the standard differential geometry.

Physically, however, we have to consider the non-vanishing
`thickness' of the coordinate lines. Then, the physical picture
has a strip of some thickness as the equatorial axis and another
strip as the axis of the `great circle'. The ` physical'
coordination of the spherical surface then refers to these `thick'
coordinate axes.

Having chosen the `physical coordination' of the spherical surface
as above, we consider a small `region' of this surface, adjacent
to the coordinate strip, as a `physical' object. Consider next
that this physical object moves. The motion of physical object is
then required to `affect' and deform the chosen adjacent physical
coordinate strip(s). A general mathematical description of this
situation is what is then required by the physical situation under
consideration.

In this situation, the principle of general relativity then
`demands' that the laws of motion of this `physical object' be
stated, and that these be the `same' mathematical statements, in
relation to `all' the permissible such physical coordinate systems
of the spherical surface.

Evidently, the mathematical formalism of the standard differential
geometry does not provide us this description since, within its
formalism, the coordinate axes of the geometry of surface, once
chosen, do not `deform' when we `move' a region of surface. Hence,
the standard differential geometry cannot faithfully implement
Einstein's principle of general relativity \footnote{As the
standard differential geometry does not do justice to Einstein's
principle of general relativity, its use through Einstein's
equations is inappropriate for physical problems  \cite{smw-utr,
smw-sars}. Einstein had, in fact, abandoned \cite{schlipp} the
formulation which he himself had proposed and had dubbed the field
equations proposed by him as ``preliminary equations' of an
attempt to test the usefulness of the ideas of general
relativity.}.

Clearly, changes that occur to the ``physical construction'' of
reference systems \cite{schlipp} must also be the part of any such
description. Then, the equality of the description of physical
phenomena must also hold incorporating any possible physical
changes to the constructions of reference systems. This situation
necessitates appropriate ``changes'' also at the conceptual
levels.

That is to say, the validity of the principle of general
relativity not only demands a mathematical framework that
incorporates the physical construction of reference systems but
also demands that we ``redefine'' various physical notions. I will
refer to this general theoretical framework \cite{smw-utr} as the
{\em Universal Theory of Relativity}.

In \cite{smw-utr}, I had discussed a mathematical framework of the
theories of measures and dynamical systems for this universal
relativity. It could however be considered to be not sufficiently
general and, hence, not entirely satisfactory. In \cite{smw-sars},
I had therefore proposed that the mathematical framework of
Category Theory \cite{cat-0, cat-1} is an appropriate basis for
the universal relativity and had discussed certain related
mathematical issues.

For the sake of wider attention, I will recall here part of that
discussion \cite{smw-sars} about relevant mathematical notions
\cite{banaschewski, macmor} and how these notions could
incorporate the ideas of universal relativity. This discussion
then essentially focuses on the aforementioned guiding principle.
Needless to say here, much more work is still needed.

For the sake of readability, I first recall some relevant
definitions and results \cite{banaschewski, macmor}.

A {\em category}, $\mathcal{C}$, consists of {\em two\/}
collections - one called the collection  $\mathcal{C}_o$ of {\em
objects\/} (denoted by capital letters A, B, C, ...) and the other
called the collection  $\mathcal{C}_A$ of {\em arrows\/} (denoted
by small letters a, b, c, ...) and {\em four operations}. First of
these operations associates with each arrow $f$ an object of
$\mathcal{C}$ called its {\em source or domain\/} (denoted by
$d_o(f)$). Second operation associates with $f$ an object of
$\mathcal{C}$ called its {\em target or co-domain\/} (denoted by
$d_1(f)$). We write $f: C \to D$ or $C \morphf D$ to indicate that
$f$ is an arrow with source $C$ and target $D$. Third operation
associates with every object $C\in\mathcal{C}_o$ an arrow, denoted
by $id_C$ and called the identity arrow of C, with source and
target as $C$. The fourth operation associates with any pair of
arrows $(f,g)$ an arrow $f\circ g$, called the composition of $f$
and $g$, such that $d_o(f)=d_1(g)$.

The above data are subject to the following four, {\em
naturally arising}, conditions: \beq d_o(1_C) &=& C = d_1(1_C) \n \\
d_o(f\circ g) &=& d_o(g), \;\;\;d_1(f\circ g) = d_1(f) \n \\
1_D\circ f &=& f, \;\;\; f\circ id_C = f \n \\ (f\circ g) \circ h
&=& f \circ (g\circ h) \n \eeq

For any two objects $C$ and $D$ of $\mathcal{C}$, the collection
of all arrows from $C$ to $D$ is called the hom-set and is denoted
by $Hom_{\mathcal{C}}(C,D)$. We will assume a fixed {\em
Universe\/} $U$ of sets and call members of $U$ {\em small sets}.
When the hom-set is a small set for any pair of objects of
$\mathcal{C}$, the category is called {\em locally small}. We will
always work with either small or locally small categories.

From $\mathcal{C}$, we form another category called the dual or
the opposite category, denoted by $\mathcal{C}^{\mathrm{op}}$,
with the same objects as in $\mathcal{C}_o$ but with directions of
all arrows and the order of all compositions of arrows in
$\mathcal{C}_A$ being {\em reversed}.

An arrow $C\morphf D$ is called a {\em monic\/} if for any $E\in
\mathcal{C}_o$ and any $g,h: E\rightrightarrows C$ in
$\mathcal{C}_1$, $f\circ g = f\circ h$ implies $g=h$, and we write
$f:C \rightarrowtail D$.

Two monic arrows $f: A\rightarrowtail D$ and $g: B \rightarrowtail
D$ are called {\em equivalent\/} if there exists an isomorphism
$h: A \isomorph B$ such that $g\circ h = f$.

A {\em sub-object of $D\in \mathcal{C}_o$\/} is an equivalence
class of monic arrows into $D$. The collection
$Sub_{\mathcal{C}}(D)$ of sub-objects of $D\in \mathcal{C}_o$ then
comes equipped with a natural partial order: $[f]\leq [g]$ if and
only if there is $h:A\to B$ with $f=g\circ h$, where the square
brackets denote the equivalence class of the monic under
consideration. The concept of a sub-object corresponds to that of
a subset of a given set or to that of a subspace of a given space.

An {\em equalizer\/} of (parallel) arrows $f,g: A\rightrightarrows
B$ is an arrow $e:E\to A$ with $f\circ e = g\circ e$ and such that
given any other arrow $u:X\to A$ with $f\circ u = g\circ u$, there
exists a {\em unique\/} arrow $v:X\to E$ with $e\circ v = u$. The
notion of an equalizer in $\mathcal{C}^{\mathrm{op}}$ corresponds
to a {\em co-equalizer\/} in $\mathcal{C}$.

An object $0$ or $0_{\mathcal{C}}$ in $\mathcal{C}$ is called as
the {\em initial\/} object if, for every other object $D$ of
$\mathcal{C}$ there exists one and only one arrow from $0$ to $D$.
An object $1$ or $1_{\mathcal{C}}$ in $\mathcal{C}$ is called as
the {\em terminal\/} object if, for every other object $D$ of
$\mathcal{C}$ there exists one and only one arrow from $D$ to $1$.

An object $X\in \mathcal{C}_o$ when equipped with arrows $\pi_1:X
\to A$ and $\pi_2:X\to B$ is called the {\em product of $A$ and
$B$\/} if and only if for any other $Y\in \mathcal{C}_o$ and any
two arrows $f:Y\to A$ and $g:Y\to B$, there exists a {\em
unique\/} arrow $h:Y\to X$ such that $\pi_1\circ h = f$ and
$\pi_2\circ h = g$.

A product in the dual category $\mathcal{C}^{\mathrm{op}}$
corresponds to what is called the {\em co-product\/} in
$\mathcal{C}$.

Consider objects $A, B, C \in \mathcal{C}_o$ with $f:B\to A$ and
$g:C\to A$. Next, consider an object $P\in \mathcal{C}_o$ with
arrows $p:P\to B$ and $q:P\to C$ such that $f\circ p = g\circ q$.
Such a collection of objects and arrows is called a commutative
square. A commutative square is a {\em pullback or
fibred-product\/} of $B, C$ over $A$, written $B\times_A C$, iff,
given any $X\in \mathcal{C}_o$ and arrows $\beta:X \to B$ and
$\gamma:X \to C$ with $f\circ\beta=g\circ\gamma$, there is a
unique arrow $\delta: X\to P$ such that $p\circ \delta =\beta$ and
$q\circ \delta = \gamma$. When pullback exists, we call $p$ as the
pullback of $g$ along $f$ or $q$ as the pullback of $f$ along $g$.

A pullback in $\mathcal{C}^{\mathrm{op}}$ corresponds to the
notion of a {\em pushout\/} in $\mathcal{C}$.

A {\em functor}, $F$, from category $\mathcal{C}$ to category
$\mathcal{D}$, is an operation which assigns to each $C\in
\mathcal{C}_o$ an object $F(C)\in \mathcal{D}_o$, and to each
$f\in \mathcal{C}_A$ an arrow $F(f)\in \mathcal{D}_A$ such that
\beq F(d_o(f)) &=& d_o(F(f)) \n \\ F(d_1(f)) &=& d_1(F(f)) \n \\
F(1_C) &=& 1_{F(C)} \n \\ F(f\circ g) &=& F(f)\circ F(g) \n \eeq
whenever $f\circ g$ is defined.

For categories $\mathcal{C}$ and $\mathcal{D}$, a functor
$F:\mathcal{C}^{\mathrm{op}} \to \mathcal{D}$ is called as {\em
contra-variant\/} from $\mathcal{C}$ to $\mathcal{D}$. A functor
$F:\mathcal{C}\to \mathcal{D}$ is called {\em full (faithful)\/}
if for any $C, C'\in \mathcal{C}_o$, the operation
$Hom_{\mathcal{C}}(C',C) \to Hom_{\mathcal{D}}(FC',FC)$, $f\mapsto
F(f)$, is {\em surjective (injective)}.

The operation $C'\mapsto Hom_{\mathcal{C}}(C',C)$ for fixed $C$
yields a contra-variant functor, called Hom-functor, from
$\mathcal{C}$ to $\mathbb{S}ets$, where $\mathbb{S}ets$ is the
category of sets. Also, $C\mapsto Hom_{\mathcal{C}}(C',C)$ for
fixed $C'$ provides a covariant Hom-functor. Functors that are
isomorphic to the Hom-functor are said to be {\em representable
functors}.

Given functors $F,G:\mathcal{C}\rightrightarrows\mathcal{D}$, a
{\em natural transformation\/} $\alpha:F\to G$ is an assignment of
each object $C\in\mathcal{C}_o$ to an arrow $\alpha_C:FC \to GC$
in $\mathcal{D}_1$, called component of $\alpha$ at $C$, such
that, for any $f:C'\to C$ in $\mathcal{C}_1$, $G(f) \circ
\alpha_{C'} = \alpha_C\circ F(f)$.

If $\alpha:F\to G$ and $\beta:G\to H$ are two natural
transformations between functors from category $\mathcal{C}$ to
category $\mathcal{D}$, a composite natural transformation,
denoted $\beta\circ\alpha$, is then naturally definable by setting
$(\beta\circ\alpha)_C =\beta_{G(C)}\circ\alpha_C$.

For fixed categories $\mathcal{C}, \mathcal{D}$, we can then
construct a {\em functor category}, denoted by
$\mathcal{D}^{\mathcal{C}}$, with its objects as functors from
$\mathcal{C}$ to $\mathcal{D}$ and its arrows as natural
transformations of these functors.

Given two functors in opposite directions, that is,
$F:\mathcal{C}\to \mathcal{D}$ and $G: \mathcal{D}\to\mathcal{C}$,
we say that $F$ is {\em left adjoint\/} to $G$, or, equivalently,
$G$ is {\em right adjoint\/} to $F$, written $F\dashv G$, when for
any $C\in\mathcal{C}_o$ and $D\in\mathcal{D}_o$, there exists a
bijective correspondence $\theta: Hom_{\mathcal{C}}(C, GD)
\isomorph Hom_{\mathcal{D}}(FC, D)$ with the arrow $f:C\to GD$
uniquely determining the arrow $h:FC\to D$ and conversely.

With the {\em adjunction\/} as above, the unique arrow $\eta_C: C
\to GFC$ with $\theta(\eta_C)=1_{F(C)}$ is called the {\em unit of
adjunction at $C$}. The dual notion is that of the {\em co-unit of
adjunction\/} which also is a unique arrow $\varepsilon_D: FGD \to
C$.

When product $A\times B$ of any pair of objects
$A,B\in\mathcal{C}_o$ exists, it provides the product functor
$\times: \mathcal{C}^{\mathbf{2}}\to \mathcal{C}$, where
$\mathbf{2}$ is the 2-object category, with only identity arrows
of those objects. The product functor is right adjoint to the {\em
diagonal functor\/} $\triangle:\mathcal{C}\to
\mathcal{C}^{\mathbf{2}}$ sending $C\in \mathcal{C}_o$ to $C\times
C$.

Let products exist in $\mathcal{C}$. Fix an object $A\in
\mathcal{C}_o$ and consider a functor $A\times -:\mathcal{C} \to
\mathcal{C}$. If a right adjoint to this functor exists, we denote
it by $(-)^A:\mathcal{C}\to\mathcal{C}$ and say that the object
$A$ is an {\em exponentiable object of $\mathcal{C}$}.

A category $\mathcal{C}$ is said to be {\em cartesian closed\/} if
it has a terminal object, binary products and if all objects of
$\mathcal{C}$ are exponentiable. For any small category
$\mathcal{C}$, the functor category
$\widehat{C}=\mathbb{S}ets^{\mathcal{C}^{\mathrm{op}}}$ is a
cartesian closed category.

Consider the functor category $\mathcal{C}^{\mathbb{J}}$ with a
fixed category $\mathcal{C}$ and a small category $\mathbb{J}$. An
object of $\mathcal{C}^{\mathbb{J}}$ is called a {\em diagram of
type $\mathbb{J}$ in $\mathcal{C}$}. Each object
$C\in\mathcal{C}_o$ determines a {\em constant diagram\/}
$\triangle_{\mathbb{J}}(C)$ having the same value $C$ for all
$j\in\mathbb{J}$. This provides a diagonal functor
$\triangle_{\mathbb{J}} :\mathcal{C}\to \mathcal{C}^{\mathbb{J}}$.

A natural transformation $\pi$ from $\triangle_{\mathbb{J}}(C)$ to
another diagram $A$ of $\mathcal{C}^{\mathbb{J}}$ therefore
consists of arrows $f_j:C\to A_j$, for each `index' $j\in
\mathbb{J}$, all such that $f_j:C\to A(j), f_k:C\to A(k),
A(u):A(j)\to A(k)$ with $u:j\to k$ form a commutative triangle for
every arrow $u\in \mathbb{J}_1$. We call $\pi$ a {\em cone $f:C\to
A$ on the diagram $A$ with vertex $C$}.

A cone $\pi:L\to A$ with vertex $L$ is {\em universal\/} when to
every other cone $f:C\to A$ corresponds a unique arrow $g:C\to L$
in $\mathcal{C}$ with $\pi_j\circ g = f_j$ for every $j\in
\mathbb{J}_o$. We call this universal cone $\pi:L\to A$ as the
{\em limit of the diagram $A$}.

If every diagram in the functor category
$\mathcal{C}^{\mathbb{J}}$ has a limit in this sense then, the
diagonal functor $\triangle_{\mathbb{J}}$ has a right adjoint -
$\lim_{\leftarrow \mathbb{J}}: \mathcal{C}^{\mathbb{J}} \to
\mathcal{C}$ - and the co-unit of this adjunction is the universal
cone, which is a natural transformation - $\pi:
\triangle_{\mathbb{J}}(L) =
\triangle_{\mathbb{J}}(\lim_{\leftarrow \mathbb{J}} A) \to A$.

The notion dual to limit is that of the {\em co-limit}. If the
co-limit of any diagram of type $\mathbb{J}$ in $\mathcal{C}$
exists then, it provides a functor $\lim_{\rightarrow \mathbb{J}}:
\mathcal{C}^{\mathbb{J}} \to \mathcal{C}$ which is left-adjoint to
$\triangle_{\mathbb{J}}:\mathcal{C}^{\mathbb{J}}\to\mathcal{C}$.

If $\mathcal{C}$, $\mathcal{D}$ are categories and
$\mathcal{C}^{\mathcal{D}}$ is the functor category, with diagrams
of type $\mathbb{J}$ in $\mathcal{C}$ existing in $\mathcal{C}$
($\mathbb{J}$ being a small category), then the same holds also
for $\mathcal{C}^{\mathcal{D}}$, and the {\em evaluation
functor\/} $(-)_D:\mathcal{C}^{\mathcal{D}} \to \mathcal{C}$, for
any $D\in\mathcal{D}_o$, preserves such limits.

Consequently, for each $D \in \mathcal{D}$, $(\lim_{\leftarrow
\mathbb{J}}A)(D) \cong \lim_{\leftarrow \mathbb{J}}A_D$ where the
limit on the left is in $\mathcal{C}^{\mathcal{D}}$ and that on
the right is in $\mathcal{C}$. Similarly, for the co-limits,
$(\lim_{\rightarrow \mathbb{J}}A)(D) \cong \lim_{\rightarrow
\mathbb{J}}A_D$.

Now, a {\em sub-object classifier\/} is a monic arrow
$\mathbf{true}: 1_{\mathcal{C}}\rightarrowtail \Omega$, where
$\Omega\in\mathcal{C}_o$, such that for every monic
$g:S\rightarrowtail X$ in $\mathcal{C}_1$, with
$S,X\in\mathcal{C}_o$, there exists a unique arrow
$\phi:X\to\Omega$, making a pullback diagram of $f:S\to
1_{\mathcal{C}}$, $g$, $\phi$ and $\mathbf{true}$. This evidently
requires the category $\mathcal{C}$ to have finite limits, {\em
ie}, to admit a terminal object $1_{\mathcal{C}}$. This
generalizes the set-theoretic notion of the characteristic
function of the subset of a set.

Then, in a category $\mathcal{C}$ with finite limits, every
sub-object is uniquely a pullback of the ``universal'' monic
$\mathbf{true}: 1_{\mathcal{C}}\rightarrowtail \Omega$. The
sub-object functor is then isomorphic to a Hom-functor, {\em ie},
$Sub_{\mathcal{C}}(X)\cong Hom_{\mathcal{C}}(X,\Omega)$. That is
to say, a sub-object functor is representable when the underlying
category $\mathcal{C}$ admits finite limits. A category is said to
be {\em well-powered\/} when $Sub_{\mathcal{C}}(X)$ is isomorphic
to a small set for all $X\in \mathcal{C}_o$.

A category $\mathcal{E}$ with (i) all finite limits and co-limits,
(ii) all objects in $\mathcal{E}_o$ being exponentiable, and (iii)
a sub-object classifier $\mathbf{true}:1_{\mathcal{E}} \to
\Omega$, is called an {\em elementary topos\/} or, simply, a {\em
topos}. A topos is, then, a complete cartesian closed category
equipped with a sub-object classifier. Examples of toposes include
$\mathbb{S}ets$, the functor category
$\mathbb{S}ets^{\mathcal{C}^{\mathrm{op}}}$ for a small category
$\mathcal{C}$, etc.

At this point, we note that the {\em Intuitionistic Propositional
Calculus\/} \footnote{Intuitionistic Logic originated with Brouwer
who insisted that any proof by contradiction be not allowed, {\em
ie}, all proofs be {\em constructive}. In this approach, {\em
all\/} functions $\mathbb{R}\to \mathbb{R}$ are {\em continuous\/}
for an appropriate description of real numbers $\mathbb{R}$.
Axiomatic foundation for Brouwer's ideas was provided by Heyting
and others.}, as formalized by Heyting, has a typical model in the
form of the {\em set of all open subsets of a topological space}.
In this model, the operations $\wedge$ and $\vee$ of the
propositional calculus correspond to intersection and union, but
the operations $\Rightarrow$ and $\neg$ have to be suitably
reinterpreted in order to yield open sets. Then, $U\Rightarrow V$
is the largest open set $W$ with $W\wedge U \subset V$, and $\neg
U$ is the interior of the complement of $U$ - the largest open set
disjoint from $U$.

Model independent characterization of this intuitionistic
propositional calculus is provided by the notion of a distributive
lattice as follows.

Any {\em partially ordered set\/} (poset) $(P, \leq)$ gives rise
to a category for which the elements of $P$ are objects and there
exists one and only one arrow from $p$ to $q$ iff $p \leq q$.
Identity arrows are forced by reflexivity and the composition of
arrows is forced by the transitivity of the partial order.

A {\em lattice\/} is a poset having, as a category, all binary
products and all binary co-products.

A poset $P$ is {\em complete\/} iff every subset of $P$ has a
lowest upper bound (lub or sup or join) and a greatest lower bound
(glb or inf or meet). A poset $P$ is then complete iff, as a
category, $P$ has all limits and all co-limits.

A {\em complete lattice $L$\/} with elements 0 and 1 with $0\leq
x\leq 1$ for all $x\in L$ is a complete poset.

A lattice $L$ with 0 and 1 is then equationally expressible as a
set with two `distinguished' elements 0, 1 and two, associative \&
commutative, binary operations $\wedge$ (meet) and $\vee$ (join)
satisfying:
\medskip

\begin{tabular}{lll} \hspace{.5in} $x\wedge x =x$,\hspace{.5in} &$x\vee x = x$
\\ \hspace{.5in} $1\wedge x =x$, &$0\vee x = x$ \\
\hspace{.5in} $x\wedge(y\vee x) =x$, &$(x\wedge y)\vee x=x$
\end{tabular} \medskip

We may then define a `lattice object' $L$ in a category
$\mathcal{C}$ with finite products as the partial order of the
lattice is recoverable from the above equations on the operations
$\wedge:L\times L\to L$, $\vee:L\times L\to L$ and
$0,1:1_{\mathcal{C}}\to L$.

A lattice $L$ is said to be {\em distributive\/} iff, for all
$x,y,z\in L$,
$$x\wedge(y\vee z)=(x\wedge y)\vee(x\wedge z)\n$$

A {\em complement\/} of $x\in L$ with 0 and 1 is then an element
$a\in L$ such that $x\wedge a=0$ and $x\vee a = 1$. In a
distributive lattice, a complement, {\em if it exists}, is
necessarily unique.

A Boolean algebra $B$ is a distributive lattice with 0 and 1 in
which every element $x$ has a complement. In $B$, $\neg\,x$ is
then interpreted as a complement. M H Stone's theorem asserts that
every boolean algebra is isomorphic to an algebra of some of the
subsets of some set.

A Heyting algebra $H$ or the Brouwerian Lattice is a poset with
all finite products, all finite co-products and is cartesian
closed as a category. Underlying lattice of any Heyting algebra
can be shown to be distributive.

An exponential $y^x$, $x,y\in H$, is usually written as
$x\Rightarrow y$ and is characterized by its adjunction: $z\leq
(x\Rightarrow y)$ if and only if $z \wedge x \leq y$.

In a Heyting algebra, we define the negation of $x$ as $\neg\,x=
(x \Rightarrow 0)$. Then, $\neg x$, is not necessarily a
complement of $x$, since $x\wedge\neg x=0$ but, not always,
$x\vee\neg x =1$. However, if a complement of $x\in H$ exists,
then it must equal $\neg x$. Also, $\neg\neg x$ need not equal $x$
in any Heyting algebra.

Moreover, the following also hold in any Heyting algebra: \medskip

\begin{tabular}{ccc} \hspace{.5in} &$x\leq \neg\neg x$,
\\ \hspace{.5in} &$x\leq y$ implies $\neg y\leq \neg x$
\\ \hspace{.5in} &$\neg x =\neg\neg\neg x$ \\
\hspace{.5in} &$\neg\neg(x\wedge y) =\neg\neg x\wedge \neg\neg y$
\end{tabular} \medskip

The `negation' is then a functor $\neg: H \to H^{\mathrm{op}}$ and
also $\neg: H^{\mathrm{op}}\to H$ which is adjoint to itself. A
Heyting algebra is then a Boolean algebra iff this adjunction is
an isomorphism. That is, a Heyting algebra is Boolean iff
$\neg\neg x = x$ for all $x\in H$ or, iff $x \vee \neg x = 1$ for
all $x\in H$. Then, quantifiers of predicate calculus like
$(\forall)$ and $(\exists)$ are also adjunctions in a Heyting
algebra setting.

In any lattice $L$ with 0 and 1, a binary operation $\Rightarrow$
satisfying, for all $x, y, z\in L$, \medskip

\begin{tabular}{ccc} \hspace{.2in} $(x\Rightarrow x) =1$
\\ \hspace{.2in} $x\wedge (x \Rightarrow y)=x\wedge y$, \hspace{.2in}
$y\wedge (x \Rightarrow y) = y$ \\
\hspace{.2in} $x\Rightarrow (y\vee z) =(x\Rightarrow y)\vee (x
\Rightarrow z)$
\end{tabular} \medskip

\noindent must be the `implication' of a Heyting algebra structure
on the lattice $L$.

In a functor category $\widehat{C}=
\mathbb{S}ets^{\mathcal{C}^{\mathrm{op}}}$ with a small category
$\mathcal{C}$, for the object $P$ of $\widehat{C}_o$, the
partially ordered set $Sub_{\widehat{C}}(P)$ of sub-objects of $P$
is, in particular, a Heyting algebra (and sometimes, a special
case, Boolean algebra).

A {\em Frame\/} is then a complete lattice $L$ in which binary
meet distributes over arbitrary joins, {\em ie}, \[ x\wedge
\bigvee S = \bigvee\left\{ x\wedge y|y\in S\right\}\] for all
$x\in L$ and $S\subseteq L$.

A {\em Frame homomorphism\/} is a map $h:L\to M$ between {\em
Frames\/} $L$ and $M$ preserving finite meets (including the top
$\mathfrak{1}$) and arbitrary joins (including the bottom $0$).

Any $a\in L$ gives rise to two new {\em Frames\/} called as the
{\em open quotient}, $\downarrow a =\{ x\in L|x\leq a\}$, and the
{\em closed quotient}, $\uparrow a = \{x\in L|x\geq a\}$,
determined by $a\in L$ with homomorphisms $L\to \downarrow a:
x\mapsto x\wedge a$ and $L\to \uparrow a: x\mapsto x\vee a$.

A particular {\em Frame}, a complete distributive lattice with 0
and 1, is then a Heyting algebra always. Also, for any topological
space $X$, the lattice of its open sets, ${\cal O}X$, forms a {\em
Frame}, being partially ordered by set inclusion.

The correspondence $X \to {\cal O}X$ is functorial with the
functor ${\cal O}: \mathbb{T}op \to \mathbb{F}rm$ being a
contravariant functor, with $\mathbb{T}op$ as the category of
topological spaces and $\mathbb{F}rm$ being the category of {\em
Frames}. Any continuous map $f: X \to Y$ determines a {\em Frame
homomorphism\/} ${\cal O}f: {\cal O}Y \to {\cal O}X$ with ${\cal
O}f(U) = f^{-\,1}(U)$ for all $U \in {\cal O}Y$.

The {\em Spectrum functor\/} in the opposite direction $\Sigma:
\mathbb{F}rm \to \mathbb{T}op$ assigns to each {\em Frame\/} $L$
its {\em spectrum\/} $\Sigma L$ that is the space of all
homomorphisms $\xi: L \to \mathbf{2}$, $\mathbf{2}$ being the
2-element lattice. Each homomorphism $\xi$ is called as a {\em
point of $L$}, and provides open sets $\Sigma_a= \left\{ \xi \in
\Sigma L | \xi{a} = 1 \right\}$ for any $a\in L$. It also assigns
to each of the {\em Frame homomorphisms\/} $h: M \to L$ a
continuous map $\Sigma h : \Sigma L \to \Sigma M$ such that
$\Sigma h(\xi) = \xi h$ with $(\Sigma h)^{-\,1}(\Sigma_a) =
\Sigma_{h(a)}$ for any $a\in M$.

The two functors ${\cal O}$ and $\Sigma$ are adjoint on the right
with the unit  of this adjunction being denoted by $\eta_L$  and
the co-unit by $\varepsilon_X$.

{\em Frames\/} for which $\eta_L$ is an isomorphism are called
{\em spatial}. ${\cal O}X$ is a {\em spatial Frame}.

Spaces for which $\varepsilon_X$ is a homomorphism are called the
{\em Sober Spaces}. Category of Sober spaces $\mathbb{S}ob$ is
dually equivalent to the full subcategory of $\mathbb{S}p\,
\mathbb{F}rm$ of {\em spatial Frames}.

Category opposite to $\mathbb{F}rm$ is called as the category
$\mathbb{L}oc$ of {\em locales\/} and it contains $\mathbb{S}ob$
as a full subcategory. Category $\mathbb{L}oc$ is often used in
the relevant mathematical literature on pointless topology which
provides a wider basis than the usual topological structure.

A {\em Frame of reals}, denoted by $\mathcal{L}(\mathbb{R})$, is a
frame generated by all ordered pairs $(p,q)$, $p,q \in
\mathbb{Q}$, subject to the following relations: \medskip

\begin{tabular}{lll} $(p,q)\wedge (r,s) = (p\wedge r,q\wedge s)$
\\ $(p,q)\vee (r,s)=(p,s)$ whenever $p\leq r< q\leq s$ \\
$(p,q)=\bigvee \left\{ (r,s)|p<r<s<q\right\}$ \\
$\mathfrak{1}=\bigvee\left\{ (p,q)|p,q\in\mathbb{Q}\right\}$
\end{tabular}
\medskip

\noindent where $\mathfrak{1}$ denotes the top of
$\mathcal{L}(\mathbb{R})$ and $\mathbb{Q}$ denotes the set of
rational numbers.

The {\em real numbers\/} are then the {\em Frame\/} homomorphisms
$\mathcal{L}(\mathbb{R})\to \mathbf{2}$, that is to say, the
points of the frame $\mathcal{L}(\mathbb{R})$ or the spectrum
$\Sigma\left(\mathcal{L}(\mathbb{R})\right)$.

A {\em continuous real-valued function\/} on an arbitrary {\em
Frame\/} $L$ is then defined as a {\em Frame\/} homomorphism
$\mathcal{L}(\mathbb{R}) \to L$. Continuous functions on $L$ are
then $L$-valued real numbers.

In the setting of pointless topology, the algebra $\mathbf{R}(L)$
of continuous real-valued functions on {\em Frame\/} $L$ is then
definable using {\em Frame\/} homomorphisms $\alpha, \beta, ...$
from $\mathcal{L}(\mathbb{R}) \to L$, with operations of this
algebra being determined by the operations of $\mathbb{Q}$ as a
lattice-ordered ring.

Next, a sheaf $F$ of sets on a topological space $X$ is a
contravariant functor $F: \mathcal{O}(X)^{\mathrm{op}} \to
\mathbf{S}ets$ such that each open covering $U=\bigcup_i, i\in I$,
of an open set $U$ of $X$ yields an equalizer diagram: \[ FU
\dashedmorph \prod_iFU_i\morphupdown \prod_{i,j} F\left(
U_i\bigcap U_j\right)\] where for $t\in FU$, $e(t) = \{t|_{U_i}
|i\in I\}$ and for a family $t_i\in FU_i$, $p(t_i) = \{ t_i|_{(U_i
\bigcap U_j)}\}$, $q(t_i)=\{t_i|_{(U_i\bigcap U_j)}\}$. Here
$\mathcal{O}(X)$ denotes the category of all open subsets of $X$,
arrows of $\mathcal{O}(X)$ being provided by set inclusions. Then,
an arrow $F\to G$ of sheaves is a natural transformation of
functors. By its definition, the category, $\mathbb{S}h(X)$, of
sheaves over a topological space  $X$ is then a full sub-category
of the functor category $\widehat{\mathcal{O}(X)}=
\mathbb{S}ets^{\mathcal{O}(X)^{\mathrm{op}}}$. The category
$\mathbb{S}h(X)$ of sheaves over the space $X$ is, in particular,
a topos.

Thus, a {\em sheaf\/} is \cite{macmor} a ``continuous set-valued''
function. A sub-sheaf of a sheaf $F$ over a space $X$ is a
sub-functor of $F$ which is itself a sheaf. Categorical co-limits
of finite limits in $\mathbb{S}h(X)$ provide {\em geometric
constructions}. These are point-wise and provide sheaves. Category
$\mathbb{S}h(X)$ of sheaves over $X$ then contains an object -
called the {\em generic set\/} - from which every other object of
$\mathbb{S}h(X)$ can be geometrically constructed \cite{macmor}.

Thus, when $X$ is a topological space, sub-sheaves of the terminal
object of $\mathbb{S}h(X)$ provide the {\em open\/} subsets of
$X$. When $X$ are Sober spaces, continuous maps between them are
functors between the corresponding categories of sheaves
preserving the geometric constructions.

Assigning, in particular, to $a\in L$ the set of all continuous
real-valued functions on $\downarrow a$ defines a sheaf on {\em
Frame\/} $L$. This sheaf is then a {\em real number object\/} in
the topos of sheaves on $L$. [Note however that this construction
does {\em not\/} extend to the category $\mathbf{F}rm$ of {\em
Frames}.]

In general, the object $\mathbf{R}_X$ of Dedekind reals in the
topos $\mathbb{S}h(X)$, where $X$ is a topological space, is
isomorphic to the sheaf of continuous real-valued functions on $X$
defined on the open sets $U$ of $X$ by $\mathbf{R}_X(U) \cong
\left\{ f:U\to \mathbb{R}|f \,\mathrm{is\,continuous} \right\}$.

$\mathbb{S}h(X)$ then provides \cite{macmor, sdg} a generalization
of a topological space - the {\em generalized space of sets\/} -
because the generic set is, in general, {\em not\/} isomorphic to
the initial object of $\mathbb{S}h(X)$, and the sub-objects of the
terminal object of $\mathbb{S}h(X)$ not necessarily its elements.
Consequently, the notion, for example, of ``continuity'' acquires
a new meaning in terms of the generic set.

Topos Theory was then developed \cite{fwlaw-01, fwlaw-02,
fwlaw-03, macmor} using a complete cartesian closed category with
a sub-object classifier - a Topos - as its fundamental structure
which is free of all set theoretic assumptions. This, Lawvere and
Tierney, axiomatization is elementary in the sense of the first
order logic with no reference to set theory.

Based on topos theoretic ideas, the differential structure built
by giving up the Law of Excluded Middle in the Euclidean setup
leads to Synthetic Differential Geometry \cite{sdg}.

Using a topos equipped with the notion of an infinitesimal
time-interval, it is then possible \cite{fwlaw-01, fwlaw-02,
fwlaw-03} to recast the newtonian framework into the topos
theoretic language. In \cite{kock-reyes}, equations of physics,
such as a wave equation and a heat equation, are also discussed
from this point of view.

Physically, the construction of a reference system must use
physical bodies. [Recall, from our earlier example, the `thick'
coordinate lines of the surface of sphere.] For its mathematical
expression, we {\em may\/} consider a physical body as a
collection of open subsets of some topological space $X$
\footnote{Objection that the coordinate axes (intervals) be
closed, compact and bounded, since such is the case with the real
line $\mathbb{R}$, does not apply here because physically
interesting quantity is only the measurable or observable
``distance'' between bodies and is definable for {\em Frames\/} as
``distance between sets''. I thank Prof B Banaschewski for raising
this issue during a short discussion.}. A physical reference
system may therefore be mathematically representable as a {\em
Frame}, a spatial {\em Frame}. Our ``guiding principle'' then
leads us, thus far, to this identification.

Generalizing on the above, the category $\mathbb{F}rm$ of {\em
Frames\/} or, equivalently, the category $\mathbb{L}oc$ of locales
could then form the underlying mathematical basis for Universal
Relativity. The category $\mathbb{S}h(X)$ would also be equally
relevant. In this framework, {\em Frame homomorphisms\/} could
then represent the occurrence of physical phenomena affecting the
reference systems.

Issues regarding the physical characteristics of material bodies
will then involve suitable {\em measures over Frames}.
Corresponding mathematical questions have not been addressed as
yet.

Furthermore, the category $\mathbb{F}rm$ is not a Topos. (There
exist many other categories which are not toposes.) Consequently,
it is unclear whether the Topos Theory could form a mathematical
basis for the universal relativity as it did \cite{fwlaw-01} for
the newtonian theoretical framework.

It may also be noted here that the work in \cite{kock-reyes}
assumes, in a topos setting, a cartesian closed category with a
commutative ring object for formulating the `wave equation' as an
ordinary differential equation with values in the vector space
(meaning $R$-module) of distributions.

But, the category $\mathbb{F}rm$ is not a topos. It is therefore
equally unclear whether the description in \cite{kock-reyes} is
any sufficiently general one.

[However, a detailed investigation of this and other issues
involved herein will surely be important to the understanding of
various implications of the universal theory of relativity.

In any case, the newtonian mathematical framework is required to
be a part of that for the universal theory of relativity.
Consequently, the mathematical framework of universal relativity
will have to `incorporate' this topos theoretic treatment of
newtonian theories.]

Still, the notion of a category is general indeed and, hence,
could itself form the basis for universal relativity. To apply
categorical notions to physical situations in universal
relativity, we then need to associate, in some manner, an
appropriate mathematical notion for `physical characteristics of
material bodies' to objects, or, equivalently, to arrows of the
category.

An important unanswered mathematical issue is then of defining an
appropriate notion of measures (or a generalization thereof)
within the basic setting of the category theory.

Notably, very general ideas from Physics and Mathematics, both,
appear to become quite identical here. Starting with the limited
formalism in \cite{smw-utr}, this ``realization'' is certainly a
significant progress. Much more work is clearly required before
the theoretical framework of the universal theory of relativity
can be mathematically represented in a satisfactory manner.

I hope to have clarified here physical principles behind the
universal theory of relativity in sufficient details and to also
have conveyed to the reader its mathematical requirements.

\acknowledgments Firstly, extensive discussions with Partha Ghosh
as well as with Gareth Amery, Sunil Maharaj and many others are
gratefully acknowledged. I thank the organizers of the South
African Mathematical Society's 48th Meeting for very good
conference arrangements and for an opportunity to present this
work. I am grateful to the School of Mathematical Sciences,
University of KwaZulu-Natal for extending to me all the support. I
am also grateful to Maga Moodley, K Komathiraj and Partha Ghosh
for giving me the enjoyable company during the beautiful drive
around the impressive Drakensburg mountain to Grahamstown and back
to the ever-enjoyable city of Durban.

\end{document}